\documentstyle[twocolumn,aps,epsf,floats]{revtex}

\newcommand{\Farka}{${\rm Farka\breve{s}ovsk\acute{y}}$}
\newcommand{\smb}{${\rm SmB_6}$}
\newcommand{\bra}[1]{\langle#1|}
\newcommand{\ket}[1]{|#1\rangle}
\newcommand{\opc}{{c}_{\vec k, \sigma}}
\newcommand{\opcd}{{c}^{\dagger}_{\vec k, \sigma}}

\begin{document}

\twocolumn[\hsize\textwidth\columnwidth\hsize\csname %
@twocolumnfalse\endcsname

\title {Hybridization effect in the Falicov-Kimball model}
\author{Minseok Park and Jongbae Hong}
\address{
	Department of Physics Education, Seoul National University, 
	Seoul 151-742, Korea}
\date{\today}
\maketitle

\begin{abstract}
The extended Falicov-Kimball model which contains hybridization between
localized level and conduction band has been studied using the exact
diagonalization method.  We obtain the single-particle density of states
and the valency from this method and analyze the behaviors of valence
transition and activation gap.  We find that the valence transition is
continuous due to hybridization, while the activation gap closes
discontinuously.  Moreover, the onset of the valence transition and the
closing point of the activation gap mismatch each other.  Our result is
well consistent with the recent experiment on \smb.
\end{abstract}
]

\section{Introduction}
During last several decades, extensive studies have been done to understand 
two-band models such as Anderson model, Kondo model\cite{Hewson}, 
and Falicov-Kimball model(FKM)\cite{Falicov}.
These models are especially useful to explain many features of lanthanides 
and actinides since these materials have broad conduction band({\it s,d-band})
and well localized valence band({\it f-band}) near Fermi level.  
These materials show some typical strong correlation effects, 
for example large effective mass, non-integer valency, and small activation gap.

We can categorize various two-band models into two groups according to
their interband interactions.  The first group includes 
hybridization between valence and conduction band.  Anderson model, periodic
Anderson model, and Kondo model are well-known examples emphasizing 
hybridization.  
Models taking interband Coulomb interaction into account fall into 
the second group. This interband Coulomb interaction is often called 
Falicov-Kimball term since they considered this interaction most crucial 
in understanding many features on lanthanides and actinides.  
FKM has been studied to explain valence transition and metal-nonmetal 
transition in non-integer 
valent rare-earth compounds$\cite{Falicov,Khomskii,Liu,Farka95a,Farka95b}$.

{\smb} and many other non-integer valent systems show phase transitions
in valence and electric property when temperature or pressure varies.
Increasing temperature makes $d$-electron more itinerant and
external or internal pressure reduces lattice constant and enhances
crystal field splitting. 
These two effects broaden conduction band and lower the energy of 
$d$-electron.  As a result, $f$-level crosses $d$-band and the system 
undergoes valence transition.  
Variation of these parameters also changes the configuration of localized 
$f$-electron and itinerant $d$-electron, which may causes metal-nonmetal 
transition.

These transitions have been explained using the hybridization gap theory
in the periodic Anderson model$\cite{Millis,Riseborough,Aeppli}$.
According to this theory, the valence transition is due to the overlap 
between $f$- and $d$-band, and the appearance of semiconducting gap is 
the result of the hybridization between two bands.
The width of the semiconductor gap is proportional to the characteristic 
energy scale, the so-called Kondo temperature ${\rm T_K}$,
and the gap disappears continuously as temperature or pressure increases.
In other words, the hybridization gap theory predicts the second order 
metal-nonmetal transition.

A generalized two-band model, the so-called extended Falicov-Kimball 
model(EFKM) considering both interband Coulomb interaction and
hybridization was proposed. 
Though much efforts have been devoted to this model for last two decades, 
no settled result on the nature of EFKM exists as far as we know. 
The most controversial point in EFKM is its valence transition behavior. 
It is well known that FKM shows discontinuous valence transition 
as pressure changes$\cite{Farka95a,Farka95b}$.  
Some authors argued that this discontinuity still remains for small 
hybridization$\cite{Avignon,Silva,Kanda,Giesekus}$.
Others, however, have reported that there is no discontinuous transition 
in EFKM$\cite{Leder,Baeck}$.
The only consensus is that the valence transition behavior in EFKM is 
so sensitive to the approximation used that only approximation-free
solution can clarify above controversy.  

In this paper, we study EFKM to show the valence transition behavior 
when hybridization exists and to elucidate metal-nonmetal transition which
has been given by Cooley {\it et al.}$\cite{Cooley95a,Cooley95b}$ recently.  
They measured for that the activation gap of {\smb}, which is a well-known 
non-integer valent material, closes discontinuously.  This result makes 
the hybridization gap theory$\cite{Millis,Riseborough,Aeppli}$
suspicious since it predicts continuous gap closing.

The transition behavior shown by Cooley {\it et al.} is strikingly 
similar to that of Mott-Hubbard transition. 
It is natural to suppose that the Coulomb interaction may play 
an important role in {\smb} and other non-integer valent materials.
The simplest model, which includes Coulomb interaction between 
localized level and conduction band, may be the Falicov-Kimball 
model(FKM)\cite{Falicov}.

Recently \Farka$\cite{Farka95a,Farka95b}$ has tried to explain 
the experimental results of Cooley {\it et al.}\cite{Cooley95a} using FKM.
He showed using the exact diagonalization method
that the Mott-Hubbard type of metal-nonmetal transition 
appears in FKM when Coulomb interaction is small.
This fact suggests that FKM may be considered as a possible microscopic model 
for non-integer valent materials such as \smb.

Since the valencem, i.e. the expectation value of f-electron number operator 
in the ground state, must be an integer in FKM,
non-integer valency is the result of statistical average of 
the mixtures of different integer valency.  
To have intermediate valency, the system must include the hybridization 
between the localized level and conduction band.

We use the Lanczos exact diagonalization method\cite{Dagotto94} to remove the
artifacts of approximation in studying this model. 
Due to the limitation of the numerical method, we treat 1-dimensional
chain of 8 lattice sites.
For this system, we first get the DOS of FKM depending on 
energy scale $\epsilon_f$, and introduce the effective gap $\Delta_{eff}$
to explain valence transition in terms of the DOS.
We find that only continuous valence transition is allowed 
when hybridization exists.
From the DOS of EFKM, we find that the activation gap shows 
sudden disappearance at a particular $\epsilon_f$ as in the case of FKM, 
while the valence changes continuously.
Moreover, the onset of the valence transition and the closing of 
activation gap show clear mismatch.
This is well consistent with the experiment on 
\smb$\cite{Cooley95a,Cooley95b}$.  

In section \ref{model}, we introduce briefly the models and method. 
We present our results on DOS, valence transition, and the activation
gap in section \ref{results}.
Section \ref{conclusion} is devoted to the summary and conclusion.

\section{Model}\label{model}
The FKM is a model explaining the properties of 2-level system
competing among binding of $f$-electron($\epsilon_f$), itinerancy of 
$d$-electron($t$), and on-site Coulomb interaction between 
$f$- and $d$-electron($U$).
Ignoring spin degree of freedom, the Hamiltonian of FKM is written as
\begin{eqnarray}
H_0 = -t\sum_{\langle i,j \rangle} (d^\dagger_i d_j + d^\dagger_j d_i)
    + U\sum_i n^f_i n^d_i + \epsilon_f\sum_i n^f_i
\label{FKM}
\end{eqnarray}
where $d^\dagger_i$($d_i$) creates(annihilates) a spinless $d$-electron at
$i$-site. $n^f_i$ and $n^d_i$ denotes the $i$-site occupation number operator
of $f$- and $d$-electron, respectively.
	
In Eq.(\ref{FKM}), we must note that $n^f=\sum_i n^f_i$ and 
$n^d=\sum_i n^d_i$ commute with the Hamiltonian.  
This commutativity implies that their expectation values, 
$\langle n^f\rangle$ and $\langle n^d \rangle$, are good quantum numbers
of this model.  Therefore $\langle n^f \rangle$ and $\langle n^d\rangle$
remain some natural numbers from the beginning to the end, and all
$\langle n^f_i \rangle$ and $\langle n^d_i\rangle$ are 0 or 1.

Due to this simplicity, there are several rigorous theorems and many numerical
results about FKM(see ref.\cite{Farka95a} and references therein).  
These are worthy of note since FKM can be applied to
various systems having $f$- and $d$-electrons.
For example, FKM has been regarded as a simplified version of 
the single band Hubbard model$\cite{Brandt91,Kennedy,Lyzawa,Vries}$, 
the crystalline formation model$\cite{Khomskii,Kennedy}$ and 
the binary alloying model\cite{Freericks}.

Ignoring the hybridization, however, could lead to an oversimplification
in describing non-integer valent systems.  In particular, it seems unreasonable
to disregard hybridization in non-integral valent systems where
the energy difference between $f$- and $d$- level is very small.
If we remind that hybridization plays a key role in the periodic Anderson 
model, it is natural to add the hybridization term to Eq.(\ref{FKM}).  
This leads us to an extended model 
\begin{eqnarray}
H = H_0 + V\sum_i (f^\dagger_i d_i + d^\dagger_i f_i)
\label{EFKM}
\end{eqnarray}
where $V$ is the transition amplitude between $f$- and $d$-level.
This is the Hamiltonian of EFKM.

Using Hartree-Fork approximation, Avignon and Ghatak\cite{Avignon} insisted
that EFKM shows discontinuous valence transition as well as continuous one
by decreasing the energy gap between two levels for a finite $f$-$d$
hybridization. Da Silva and Falicov\cite{Silva} also reported that 
they found first and second order valence transitions in EFKM 
using Green's function method with mean field approximation and 
zero conduction band width. 
According to their result, discontinuous valence transition appears 
when $U/V \geq 2.5$.  Kanda {\it et al.}\cite{Kanda} showed the same result
with Da Silva and Falicov\cite{Silva} within Hartree-Fork approximation 
but non-zero conduction band width.  
Recently, Giesekus and Falicov\cite{Giesekus} also reported using Hartree-Fork 
approximation that EFKM has first order valence transition for
particular parameter range.

However, there are some contradictory results to those mentioned above.
Using Hartree-Fork approximation and taking the periodicity of 
the system into account, Leder\cite{Leder} argued that
there is no discontinuity in valence transition in EFKM.
Baeck and Czycholl\cite{Baeck} also reported using different decoupling method 
that first order valence transition does not exist in EFKM.

Depending on the approximation used, some of results show discontinuity in
valence transition but others do show only continuous transition.  This is
why we use the Lanczos exact diagonalization method to get rid of the
effect of approximation.
At first, we use the modified Lanczos method to get the ground 
state\cite{Dagotto85}.  Then using the Lanczos method, we investigate 
the ground state valency $\bra{\psi_g} n_f \ket{\psi_g}$, and the DOS
\begin{eqnarray}
D_{\rm o}(\omega)&=& \sum_{\vec k,n} \left|
       \bra{\psi^{N-1}_n}\opc\ket{\psi^N_g}
        \right|^2 \delta(\omega-E^N_0+E^{N-1}_n) \nonumber\\
    &=& \sum_{\vec k} S_{\rm o}(\vec k,\omega) \label{s_pes_def}
\end{eqnarray}
and
\begin{eqnarray}
		D_{\rm e}(\omega)&=& \sum_{\vec k,n} \left|
        \bra{\psi^{N+1}_n}\opcd\ket{\psi^N_g}
        \right|^2 \delta(\omega+E^N_0-E^{N+1}_n) \nonumber \\
    &=& \sum_{\vec k} S_{\rm e}(\vec k,\omega)~, \label{s_ipes_def}
\end{eqnarray}
where subscript 'o' and 'e' mean 'occupied' and 'empty', respectively. 
$S(\vec k,\omega)$, the spectral function, means the partial DOS of momentum
$\vec k$.  $S(\vec k,\omega)$ can be compared with the angle resolved
photoemission spectroscopy data.

\section{Results}\label{results}
\begin{figure}[ht]
\epsfxsize=8.5cm
\centerline{ \epsffile{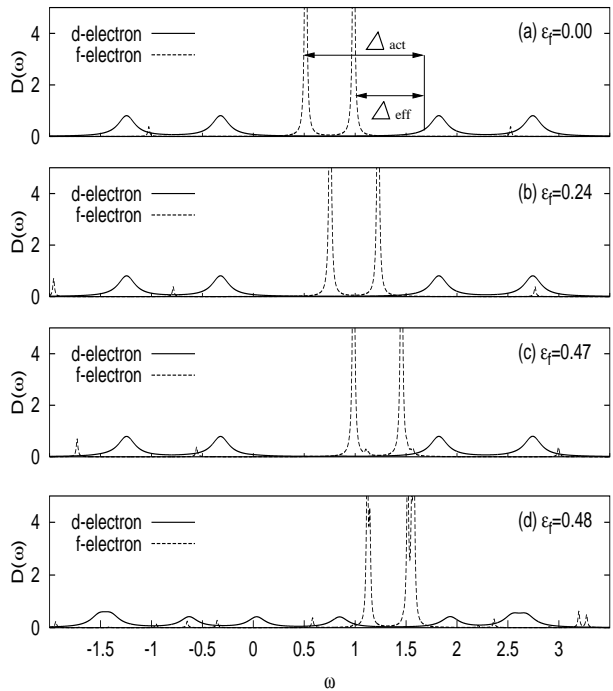}}
\caption{ 
	The DOS for FKM with $V$=0 and $U$=1.5.
	Fermi surfaces are located in between two large $f$-peaks in all cases.
}
\label{simple_dos}
\end{figure}

The $f$- and $d$-electron DOS of FKM depending on $\epsilon_f$ are shown in 
Fig.\ref{simple_dos}.  
We put in the figure the activation gap $\Delta_{act}$ corresponding to 
the energy gap between filled $f$-band and empty $d$-band
and the effective gap $\Delta_{eff}$ corresponding to the energy gap
between empty $f$-band and empty $d$-band.

\begin{figure}[ht]
\epsfxsize=8.5cm
\centerline{ \epsffile{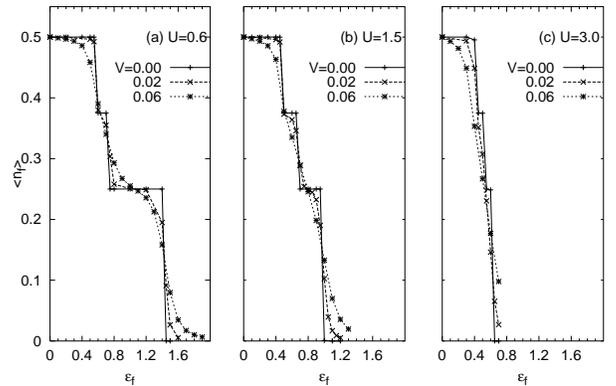}}
\caption{ 
	$\langle n_f \rangle$ depending on $\epsilon_f$ for $U$=0.6(a),
	1.5(b), and 3.0(c).  Hybridization effect is shown in each case
	with $V$=0, 0.02, and 0.06.  Hybridization effect is more
	clear within large $U$ or large $\epsilon_f$ region.
}
\label{vt}
\end{figure}

The ground state electron configuration in real space for 
$\epsilon_f \leq 0.47$ 
is the alternating type such as {\it -f-d-f-d-}.  
There is no sign of change in electron configuration from $\epsilon_f$=0(a), 
to $\epsilon_f$=0.47(c).   The only change in DOS is shifting of $f$-DOS,
to high-frequency side.
On the contrary, Fig.\ref{simple_dos} (d) is quite different from 
Fig.\ref{simple_dos} (a)-(c), which implies that sudden change takes place
in between $\epsilon_f$=0.47 and $\epsilon_f$=0.48.
This is clear from Fig.\ref{vt} which shows $f$-electron valency 
$\langle n_f \rangle$ depending on $\epsilon_f$.
In Fig.\ref{vt} (b) for $U$=1.5, valence transition occurs at 
$0.47 < \epsilon_f < 0.48$, where $\Delta_{eff}$=0.

Therefore it is clear that $\Delta_{eff}$ is the effective gap 
governing the valence transition in FKM.
The difference between $\Delta_{act}$ and $\Delta_{eff}$ is resulted from 
onsite electron-hole binding and proportional to the onsite interband 
Coulomb interaction strength $U$.
This interpretation is well consistent with the original idea of 
Falicov and Kimball\cite{Falicov} about the origin of 
discontinuous valence transition in two band system.
Thus $\Delta_{eff}$ corresponds to the {\it shrinking gap} 
\begin{eqnarray}
\Delta_{eff} = \Delta_{act} - 2U(1-\langle n_f \rangle)
\label{eq_delta}
\end{eqnarray}
suggested by Falicov and Kimball\cite{Falicov}.

Fig.\ref{vt} shows the type of valence transition.
For $V$=0 the $f$-electron valency shows steps as $\epsilon_f$ increases,
which is the well-known discontinuous valence transition in FKM.
The hybridization effect in the valence transition is also clearly shown
in Fig.\ref{vt}.  Within all parameter range, all edges of $V$=0
disappear for any small hybridization.  This disappearance clarifies 
continuous valence transition in all $U$ for $V \neq 0$.
It is interesting to see that the hybridization effect is more prominent 
in large $U$ and large $\epsilon_f$.
This is consistent with the reduction of $\Delta_{eff}$, which is 
corresponds to the length of plateau in Fig.\ref{vt}, 
as $U$ or $\epsilon_f$ increases.
Eq.(\ref{eq_delta}) clearly explains these behaviors of $\Delta_{eff}$.

\begin{figure}[ht]
\epsfxsize=8.5cm
\centerline{ \epsffile{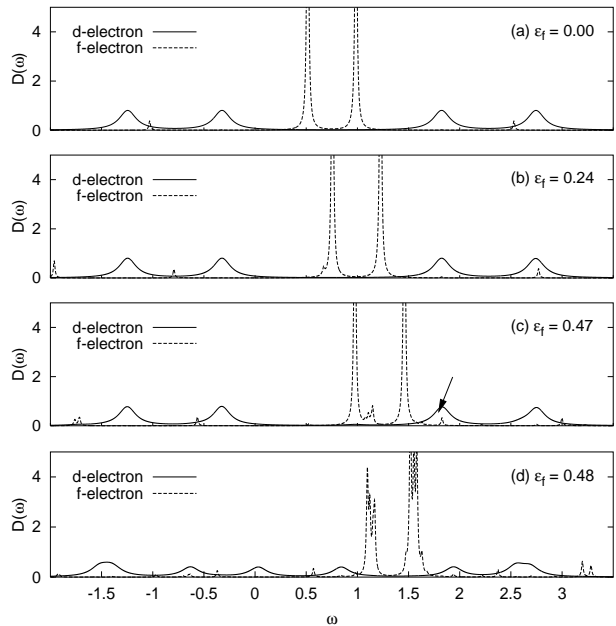}}
\caption{ 
	DOS for $U$=1.5 and finite hybridization $V$=0.02.  
	Fermi surfaces are located between two large $f$-peaks in all cases.
	Arrow in (c) indicates prominent difference between $V$=0.02 and
	$V$=0 case which is shown in Fig.\ref{simple_dos}.
}
\label{dos_with_V}
\end{figure}

The continuous transition can be confirmed via analyzing DOS.
We present the change of DOS with $\epsilon_f$ for a finite hybridization 
$V$=0.02 in Fig.\ref{dos_with_V}.
Comparing Fig.\ref{dos_with_V} with Fig.\ref{simple_dos}, 
we find that small $f$-peak indicated by arrow in (c) grows 
as we increase $\epsilon_f$ from 0 to 0.47.
This new creation of $f$-peak make $\Delta_{eff}$=0.
Therefore $\Delta_{eff}$ is not defined in the hybridization-induced
valence transition or $\Delta_{eff}$ is always zero for $V \neq 0$, 
and this means the transition is continuous.
As this $f$-peak grows, valence electrons can transfer more easily 
to conduction band and we observe continuous change of valency as a result.

\begin{figure}[ht]
\epsfxsize=8.5cm
\centerline{ \epsffile{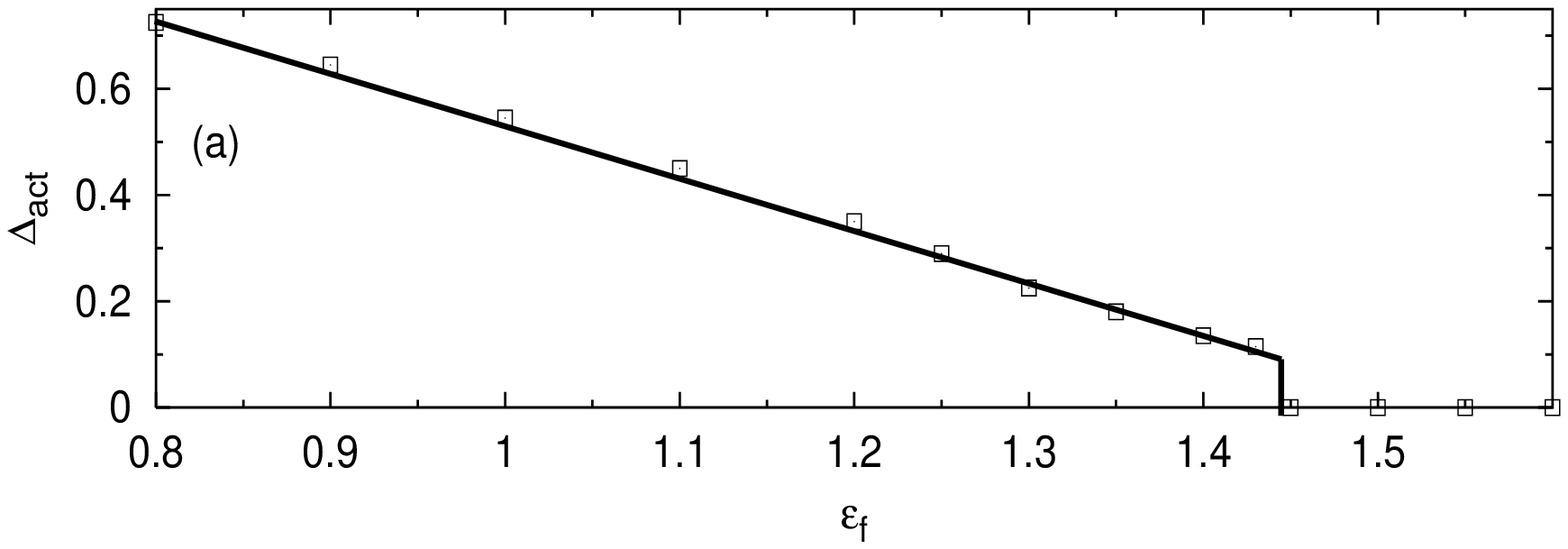}}
\vspace{-3mm}
\epsfxsize=8.5cm
\centerline{ \epsffile{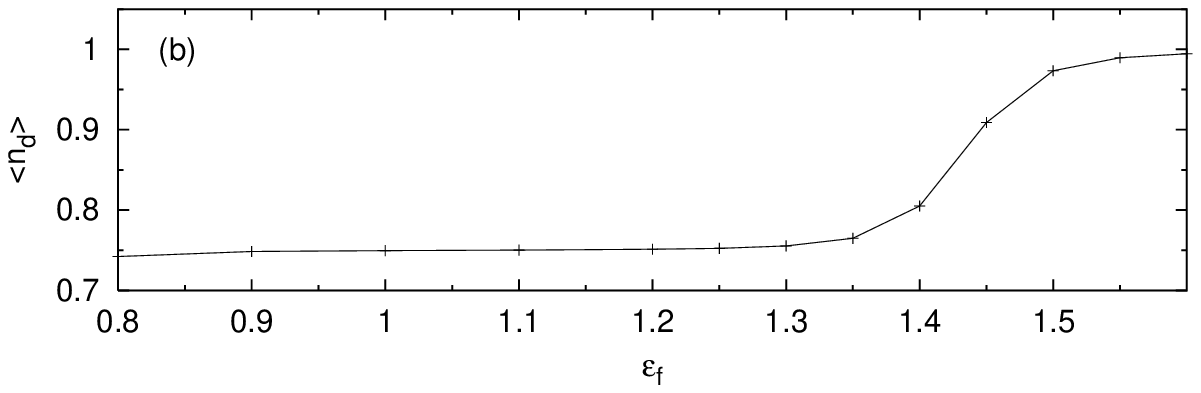}}
\caption{ 
	Results on (a) activation gap $\Delta_{act}$ and 
	(b) conduction electron density $\langle n_d \rangle$ 
	depending on $\epsilon_f$.  $V$=0.02 and $U$=0.6.
	Line of (a) is drawn for eye guide.
	Our results show the sudden closing of $\Delta_{act}$
	and the mismatching between the $\langle n_d \rangle$ increasing
	and $\Delta_{act}$ closing.
	These are well consistent with recent experimental results.
}
\label{gap}
\end{figure}

We now study the transport property of the system in terms of 
the activation gap $\Delta_{act}$ which has been measured by 
Cooley {\it et al.}\cite{Cooley95a}.
We calculate the activation gap $\Delta_{act}$ from DOS.
We find that $\Delta_{act}$ decreases linearly as $\epsilon_f$ increases
in the constant valence regions which correspond to plateaus in Fig.\ref{vt}.
This is quite natural since the effect of $\epsilon_f$ on DOS
is shifting of the $f$-DOS with $d$-DOS as rigid background.
Calculation has been done for $U$=0.6 and $V$=0.02 and the result 
is shown in Fig.\ref{gap}.
$U$=0.6 is the value for which {\Farka} has shown metal-nonmetal transition 
for $V$=0\cite{Farka95b}, 

Our result demonstrates that the activation gap closes not at 
the onset of valence transition but at the end of valence transition.
This is the well-known mismatching showing in the experiment of
Cooley {\it et al}.

\section{Conclusion}\label{conclusion}
In this paper, we consider a generalized two band model, the so-called
extended Falicov-Kimball model which includes both the Coulomb repulsion 
and hybridization between valence band and conduction band.
The behaviors of valence transition and metal-nonmetal transition in EFKM
have been studied using the Lanczos method.

First, we give a remark on the energy gaps.
The activation gap $\Delta_{act}$ governs transport properties
of the system, while the effective gap $\Delta_{eff}$ governs 
the valence transition. 
The latter is the effective gap suggested by 
Falicov and Kimball\cite{Falicov} and 
determines the condition of valence transition in FKM.
Intrasite electron-hole pairing causes decreasing of 
$\Delta_{act}-\Delta_{eff}$.  This is known as gap shrinking by 
Falicov and Kimball\cite{Falicov}.

Secondly, we find that the valence transition becomes continuous even for 
small hybridization.
The effect of hybridization in valence transition is determined
by the competition between $\Delta_{eff}$ and $V$.
For a fixed $V$, the hybridization effect becomes more prominent 
as $U$ or $\epsilon_f$ increases.
Fig.\ref{simple_dos} clearly shows the increasing of $U$ or $\epsilon_f$ 
causes smoother transition.
The DOS of Fig.\ref{dos_with_V} shows that hybridization makes 
the valence transition continuous by showing the growing of small 
$f$-peak as $\epsilon_f$ increases, which makes the effective gap vanishes.
Continuous growth of this $f$-peak reflects continuous valence transition
when $V \neq 0$.

Finally, we show that the exotic metal-nonmetal transition behavior of 
\smb\cite{Cooley95a} can be explained by EFKM.
Due to finite hybridization the valence transition takes place continuously,
while $\Delta_{act}$ decreases linearly as $\epsilon_f$ increases 
until the system meets new valence phase. 
$\Delta_{act}$ closes suddenly at the point of valence change.
These features of EFKM are well consistent with Cooley {\it et al.}'s 
experimental result on \smb.

\end{document}